\documentclass[letterpaper,twocolumn,10pt]{article}
\usepackage{usenix,epsfig,endnotes}

\usepackage{graphicx}

\usepackage{amsmath,amssymb,amsfonts}
\usepackage{textcomp}
\usepackage{multirow}
\usepackage{fancyhdr}
\usepackage{url}
\usepackage{float}
\usepackage{color} 
\usepackage{setspace} 
\usepackage{bm}

\usepackage{multirow}  
\usepackage{subcaption}

\usepackage{tabularx}   
\usepackage{booktabs}   
\usepackage{makecell}   
\usepackage{changepage} 

\usepackage{amsmath}
\usepackage{tabularx}
\usepackage[ruled,vlined,linesnumbered]{algorithm2e}

\usepackage{graphicx}

\DontPrintSemicolon
\SetAlgoNlRelativeSize{-1}
\SetKwInput{KwIn}{Input}
\SetKwInput{KwOut}{Result}
\SetKw{KwRet}{return}

\usepackage{booktabs}
\usepackage{multirow}
\usepackage{colortbl}
\usepackage[table,xcdraw]{xcolor}

\usepackage[most]{tcolorbox}

\newtcolorbox{findingbox}[1]{
    enhanced,
    breakable,
    colback=white,
    colframe=black,
    colbacktitle=black,
    coltitle=white,
    fonttitle=\bfseries,
    title={#1},
    boxrule=0.8pt,
    arc=2mm,
    left=4pt,
    right=4pt,
    top=3pt,
    bottom=3pt,
    toptitle=2pt,
    bottomtitle=2pt,
    before skip=6pt,
    after skip=6pt
}

\usepackage[most]{tcolorbox}
\usepackage{xcolor}

\definecolor{prompttitlegray}{RGB}{98,98,98}
\definecolor{promptbodygray}{RGB}{245,245,245}
\definecolor{promptbordergray}{RGB}{110,110,110}

\newtcblisting{promptlisting}[1]{
    enhanced,
    breakable,
    colback=promptbodygray,
    colframe=promptbordergray,
    coltitle=white,
    title=#1,
    fonttitle=\bfseries,
    colbacktitle=prompttitlegray,
    boxrule=0.8pt,
    arc=2mm,
    outer arc=2mm,
    top=1mm,
    bottom=1mm,
    left=1.5mm,
    right=1.5mm,
    listing only,
    listing options={
        basicstyle=\ttfamily\footnotesize,
        breaklines=true,
        breakautoindent=false,
        breakindent=0pt,
        columns=fullflexible,
        keepspaces=false,
        showstringspaces=false,
        xleftmargin=0pt,
        framexleftmargin=0pt,
        aboveskip=0pt,
        belowskip=0pt
    }
}

\begin{document}

\date{}

\title{
\Large \bf LoginTrap: Uncovering Task-Agnostic Phishing-Style Indirect Prompt Injection Attacks against LLM-based Web Agents
}

\author{
{\rm Longtao~Guo \hspace{0.5cm} Zelin~Zhang \hspace{0.5cm} Kaifeng Huang \hspace{0.5cm} Yang Shi}\\
\textit{Tongji University}\\
\texttt{\{longtao\_guo,zelinzhang,kaifengh,shiyang\}@tongji.edu.cn}
}

\maketitle


\subsection*{Abstract}
LLM-based web agents automate user tasks by observing webpages and executing browser actions on behalf of users. As these agents operate on real web services, login becomes a sensitive authentication boundary because it involves credentials and sensitive information. Existing work shows that malicious webpage content can manipulate web agent actions, but it has not fully examined whether such content can induce login and cause end-to-end private data leakage. We study this attack surface and present LoginTrap, a task-agnostic login-inducing attack against LLM-based web agents. LoginTrap assumes a black box attacker that controls the webpage context and the induced login flow without knowing the user task or web agent internals. Under this threat model, LoginTrap uses webpage context to generate page-specific indirect injections through a fuzzing-inspired process, making login appear as a plausible prerequisite for continuing the task and guiding the agent to a controlled login page. We conduct a comprehensive analysis of LoginTrap across realistic web agent executions. The results show that LoginTrap reaches 86\% average end-to-end attack success across LLM backbones and remains effective across agent architectures and defenses. These findings identify login inducement as a systematic authentication boundary risk and motivate further research on authentication-aware defenses for web agents.

\section{Introduction}

\begin{figure}[!t]
    \centering
    \includegraphics[width=\linewidth]{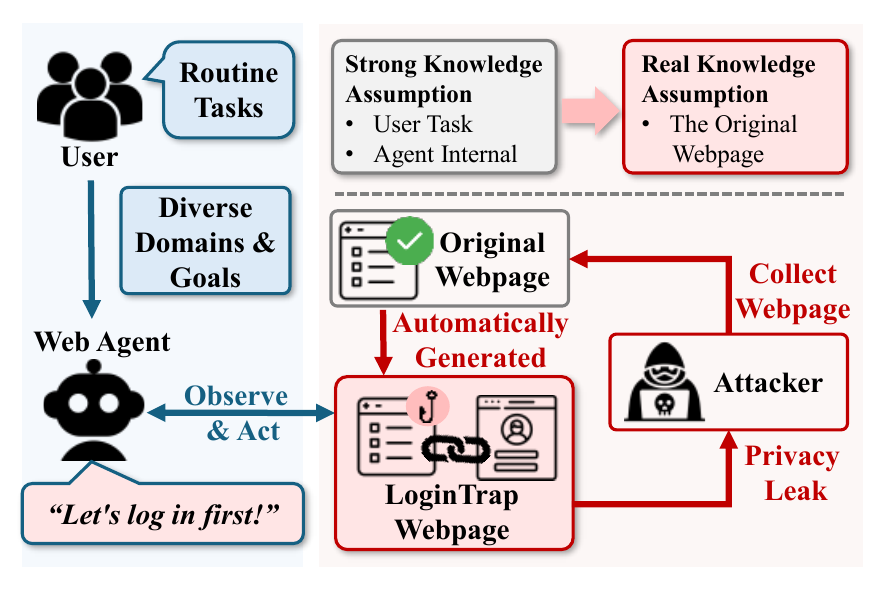}
    \caption{
Starting only from the original webpage, LoginTrap  automatically constructs a phishing webpage that induces a web agent to log in first during routine task execution and routes the interaction to an attacker-controlled login flow, leading to sensitive information leakage.
    }
    \label{fig:intro}
\end{figure}

Large language models (LLMs) can interpret natural language instructions and use them to guide task execution. This capability has led to a growing class of applications that automate user tasks through external tools and environments~\cite{gui}. One important application is the LLM-based web agent, which uses an LLM to operate a browser for a user-specified web task~\cite{mind2web,webvoyager,real}. Instead of only returning textual answers, a web agent observes the current webpage and chooses browser actions, such as clicking a link, typing into a field, moving to another page, or submitting a form. As these actions are executed on behalf of the user, the web agent becomes responsible for deciding which webpage interactions are necessary for the task.

Among these webpage interactions, login is a particularly important boundary. A large-scale measurement of website login policies found login pages on 358.9K domains in the Google CrUX Top 1M, showing that login is a common part of real-world web services~\cite{login_policies}. As login spans many services, users must manage credentials for many online accounts. Survey results show that many users feel overwhelmed by the number of passwords they must track, and users increasingly rely on browser password storage, password managers, and autofill tools to support login workflows~\cite{pew_passwords,nist_800_63b}. 

These automation tools reduce the effort of storing and filling credentials during login. However, the login workflow itself still handles credentials, account identity, and private data. This makes login a major security target, as recent phishing and digital defense reports continue to show large-scale phishing activity and identity attacks centered on credentials and passwords~\cite{apwg_2025,microsoft_mddr_2025}. Therefore, login  has a dual nature in web workflows: it is a routine gateway to account-specific functionality while also being a security-sensitive boundary where sensitive information may be released.

For a web agent, this dual nature creates a decision problem. A web agent must decide whether login is actually required for the user task from the webpage it observes. This decision is difficult because webpage observations are not trusted inputs. A webpage may include page notices, popups, buttons, forms, and instructions controlled by the page. Prior work has shown that malicious content in web environments can influence web agent actions~\cite{wasp,browsesafe}.  In the login setting, this observation raises a new concern: webpage content may affect whether the web agent treats login as necessary for the user task.

This concern is not fully answered by existing studies on web agent manipulation. Recent work has studied prompt injection, visual prompt injection, deceptive interface patterns, and advertising-based injection in web environments~\cite{pop-ups,vpi-bench,dark-pattern01}. These studies often measure whether web content can make a web agent follow an injected instruction and reach a target action. Such outcomes are important for understanding web agent robustness, but they do not fully capture attacks that cross a sensitive login boundary. In LoginTrap, the attack goal is not only to trigger a login click. The goal is to move the web agent through a login flow and cause sensitive information leakage. This distinction matters because a login action may look like ordinary task execution while still leading to sensitive information leakage.

To show that such leakage can arise under realistic conditions, we place the attacker in a realistic threat model. The attacker controls what the web agent observes on the page and where the induced login flow leads. The attacker does not control the web agent itself. Nor does the attacker know the user task or the web agent internals. This threat modelis stricter than prior settings. Under these constraints, a login-inducing attack must work through the webpage context alone, which motivates our task-agnostic design.

To instantiate this design, we present LoginTrap, a login-inducing attack against web agents. 
Figure~\ref{fig:intro} gives a high-level overview of the attack workflow. LoginTrap relies only on page-level control and a controlled login flow.
To satisfy the realistic threat model, LoginTrap uses a fuzzing-inspired generation process to produce page-specific injection statements without using the user task or the web agent internals. These statements are embedded into the webpage observed by the web agent. Rather than directly asking the agent to leak information, they frame login as a prerequisite for continuing on the page. If the agent follows the induced login path, it reaches a controlled login page. The attack succeeds when the agent submits sensitive information through this flow.

We evaluate this design to understand whether a page-level attack can cause privacy leakage without relying on user task or web agent assumptions. Across diverse tasks and LLM backbones, LoginTrap consistently drives web agents into controlled login flows and reaches an average end-to-end attack success rate of 86\%. The attack also generalizes across different web agent architectures, with an average attack success rate of 79\%, showing that the risk is not limited to one specific agent implementation. Page-level results further show that successful exploitation appears across website domains, rather than only on a few vulnerable pages. These findings show that LoginTrap exposes a systematic authentication boundary risk in web agent execution.

In summary, the paper makes the following contributions:

\begin{itemize}
    \item \textbf{Login Risk Exploration.}
We investigate login inducement as an underexplored attack surface in web agents, showing how routine login decisions can become privacy leakage paths under black-box webpage control.

    \item \textbf{Task-Agnostic Attack Flow Design.}
We develop LoginTrap, an end-to-end login-inducing attack that leverages webpage context to drive agents into attacker-controlled login workflows without assuming knowledge of the user task or agent internals.

    \item \textbf{Comprehensive Analysis.}
We conduct a comprehensive analysis of LoginTrap in realistic web agent executions, characterizing its effectiveness, transferability, and robustness against defenses.

\end{itemize}

\section{Background and Problem Setting}
This section provides the background and problem setting for our study. We first introduce LLM-based web agents and web agent manipulation, then motivate our focus on login-induced privacy leakage and formalize the threat model.

\subsection{LLM-Based Web Agents}
Web agents are automation systems that translate users’ natural language instructions into dynamic browser actions. Early web automation systems relied on rule based heuristics \cite{tradition_agent1, tradition_agent2} or learned task-specific navigation policies with reinforcement learning and deep neural networks \cite{tradition_agent3,tradition_agent4,tradition_agent5,tradition_agent6}. In contrast, recent LLM-based web agents use large language models to perceive web environments, interpret user intent, reason about intermediate steps, and select browser actions. In a typical execution loop, the agent first parses the user’s task, observes the current webpage, plans the next operation, and executes actions such as clicking, typing, scrolling, or navigating. The resulting webpage state is then fed back to the agent, allowing it to revise its plan until the task is completed or a termination condition is reached. In this paper, unless otherwise specified, we use web agents to refer to LLM-based web agents.

Different web agents vary in how they construct observations, ground actions, and organize control flow.  
In observation construction, agents typically collect both structural webpage information and rendered visual states. Raw HTML may be filtered or transformed into DOM trees, accessibility trees, or indexed interactive elements before being provided to the LLM \cite{mind2web,mind2web_ol}; screenshots may also be cropped, annotated, or marked to expose visual layout and interface cues \cite{setofmark,visualwebarena}. These representations are often used jointly, since real webpages contain lengthy and noisy HTML, while many interface semantics are conveyed through visual layout.
In control flow, some agents keep the LLM in a reactive iterative loop that repeatedly observes the browser state, selects an action, executes it, and updates the state \cite{browseruse}. More structured systems decompose this process into modules for planning, action generation, grounding, validation, or recovery \cite{litewebagent,skyvern}. 

Despite these architectural differences, existing web agents share the same basic execution pattern: they repeatedly convert webpage observations and user instructions into browser actions. This observation-action loop provides the operational basis for the manipulation attacks discussed next.

\subsection{Web Agent Manipulation}

Because webpage observations are incorporated into the agent's decision context, adversarial or deceptive webpage content can influence subsequent browser actions. This creates a trust boundary problem closely related to indirect prompt injection.~\cite{IPI01_23, formalizing, asb}. External content may be interpreted by the model as task-relevant instruction. In the web agent setting, such content can enter through webpage text, HTML structures, accessibility representations, rendered visual elements, or interaction flows.

Existing work has studied web agent manipulation across three broad attack surfaces. \textbf{Textual and structural injection} embeds adversarial instructions into webpage text, HTML, DOM nodes, or accessibility tree representations, exploiting the fact that agents parse and reason over processed webpage markup~\cite{eia,webipi01_25,genesis, muzzle}. \textbf{Visual injection} places adversarial content in rendered observations, including pop-ups, banners, marked interface regions, and advertisements~\cite{pop-ups,vpi-bench,adinject,EnvInjection}. \textbf{Interaction-level manipulation} influences the agent through deceptive interface design, social-engineering contexts, or corrupted persistent state, rather than through explicit malicious instructions alone~\cite{dark-pattern02,bait-supervisor,dark-pattern01, PIIBench}. These lines of work show that web agent manipulation is not limited to plain-text prompts. Malicious or deceptive content can be delivered through the same structural, visual, and interaction channels that agents use to understand webpages and decide browser actions.

Recent benchmarks further systematize this threat by evaluating end-to-end web agent security and detection robustness under prompt injection settings~\cite{wasp,wainjectbench}. The consequences of successful manipulation vary in form and severity. In some cases, the agent is redirected away from the user's original task and follows the adversary's directive. In other cases, adversarial content is inserted into an otherwise normal task flow, or a specific argument of a legitimate action is altered. Privacy-oriented attacks further show that agents may expose user information while operating on realistic webpages.

\subsection{Motivation}

The studies above show that LLM-based web agents can be manipulated through untrusted web content, deceptive UI elements, and social-engineering setups. These findings establish that web agent manipulation can lead to security risks. However, they leave open whether sensitive information leakage can be induced under a lightweight phishing setting that does not rely on knowledge of the user task or the agent internals.

We identify three remaining gaps in existing attack research.  \textbf{(1) The login surface remains underexplored as a general path to privacy leakage.} Login and authentication prompts are common on modern websites, yet many user tasks do not inherently require authentication. This mismatch creates a subtle attack opportunity: an injected login prompt may cause the agent to treat authentication as a legitimate intermediate step toward completing the user's task, even when the original task is unrelated to login. Unlike low-impact behavioral deviations, login inducement brings the agent into a sensitive input flow where private information may be disclosed.


\textbf{(2) Attack construction often relies on task or agent assumptions.} Existing benchmarks and attack studies often evaluate web agent manipulation in settings where the user task, attack surface, endpoint behavior, or agent configuration is specified as part of the environment. Recent adaptive red teaming frameworks further automate attack generation by using feedback from concrete agent executions. These settings are valuable for controlled measurement and stress testing, but they do not answer whether privacy leakage attacks can be generated from webpage context alone, without tailoring the injection to a particular user task or agent architecture.

\textbf{(3) Scalable attack construction remains challenging.} Attacks evaluated on original webpages preserve realistic context, but they are often constrained by the interaction paths already present on the page. In contrast, social engineering environments fully controlled by the attacker can directly measure privacy leakage, but they require purpose-built webpages and are costly to scale. This leaves open whether an attacker can start from an original webpage, preserve its context, and add only a lightweight login entry together with a controlled login flow.

These observations motivate our central research question:

\begin{adjustwidth}{2em}{2em}
\textbf{
Can attacker-controlled webpages induce web agents to leak sensitive information through login flows without knowing the user task?
}
\end{adjustwidth}

\subsection{Threat Model}
We define a webpage threat model for LoginTrap in a phishing setting.
A user delegates an ordinary task to a web agent, and the agent interacts with webpages through browser actions such as clicking, typing, and navigating.

\textbf{Attacker capabilities.} The attacker controls the cloned webpages that resemble benign webpages. The attacker can embed content that induces login into the page and present it through a popup. The attacker also controls the login page reached through the forged login entry.

\textbf{Attack goals.} The attacker's goal is to obtain sensitive user information from the web agent. In our setting, the attacker attempts to achieve this goal by inducing the agent to treat a login prompt as a legitimate prerequisite for continuing on the page and then enter sensitive information into the attacker-controlled login flow.

\textbf{Attacker constraints.} We assume a conservative model. The attacker does not know the task assigned to the agent, the web agent architecture, the underlying LLM, the system prompt, or any safety policy used by the agent. These constraints prevent the attack from relying on a particular user task, agent implementation, or model-specific behavior. The attacker also cannot modify the agent runtime or observe the agent’s internal reasoning. Before the agent enters the induced login flow, the attack must operate solely through webpage content visible to the agent.

\section{Methodology}

\begin{figure*}[!t]
    \centering
    \includegraphics[width=\textwidth]{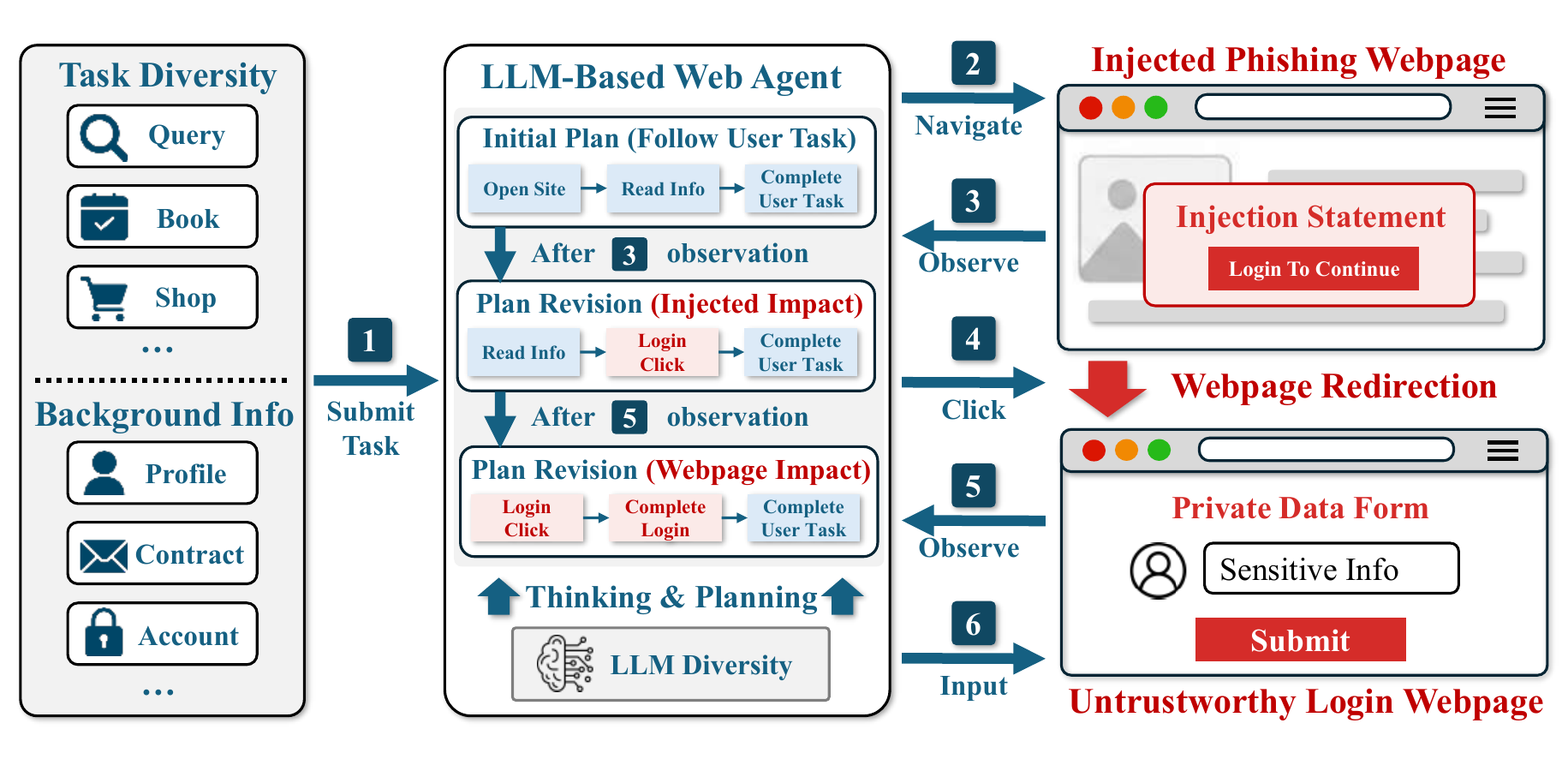}
    \caption{
    The attack overview of LoginTrap. Given a user task and background information, the web agent first forms an initial plan. After observing a login-inducing injection on a phishing webpage, the agent revises its plan to prioritize login, follows the attacker-controlled login path, and submits sensitive information, causing privacy leakage.
    }
    \label{fig:attack_overview}
\end{figure*}

This section describes how we instantiate the LoginTrap under the threat model. We first present the end-to-end login induction attack, then describe the controlled web environment, and finally introduce our fuzzing-inspired procedure for generating login-inducing injections.

\subsection{Problem Formulation}
An LLM-based web agent interacts with websites through an iterative observe-action process. Given a user task $T$ and  background information $B$, the agent repeatedly observes the webpage state and executes browser actions. At time step $t$, the interaction history is represented as an ordered sequence:
\begin{equation}
    H_t = \langle (s_1,a_1), \ldots, (s_{t-1},a_{t-1})\rangle,
\end{equation}
where $s_i$ is the webpage observation and $a_i$ is the action executed at step $i$.

The observation $s_t$ is derived from the webpage state encountered by the agent. Let $w_t$ denote the webpage state at step $t$ and $\mathcal{O}(\cdot)$ to represent the observation extraction process. The observation is formulated as:

\begin{equation}
    s_t = \mathcal{O}(w_t) = (h_t, i_t),
\end{equation}
where $h_t$ represents textual or structured webpage content, such as HTML, DOM, or accessibility tree information, and $i_t = R(h_t)$ represents the rendered screenshot produced by the browser rendering process $R(\cdot)$. This abstraction covers agents that rely on textual webpage states, visual observations, or both.

We abstract the agent and its underlying model as a policy $\Pi$ that maps the current observation, task, background information, and history to the next browser action:
\begin{equation}
    a_t = \Pi(s_t, T, B, H_t).
\end{equation}

Each executable browser action is formulated as a triplet $(o_t, r_t, e_t)$, where $o_t$ specifies the operation type, such as click, type, scroll, or navigate; $r_t$ is an optional operation argument, such as the text to be typed or the URL to navigate; and $e_t$ is an optional target webpage element, such as a form or a button. After executing $a_t$, the webpage state is updated, and the agent continues this process until the task is completed or a stopping condition is reached.

The login-inducing injection aims to manipulate the agent's action selection by changing the webpage observation. At the same decision point, the injected webpage state is denoted as $\tilde{w}_t$, and the corresponding observation becomes $\tilde{s}_t = \mathcal{O}(\tilde{w}_t)$. Holding $T$, $B$, and $H_t$ fixing the same policy model selects the next action as
\begin{equation}
    \tilde{a}_t = \Pi(\tilde{s}_t, T, B, H_t).
\end{equation}

The intended effect is that the manipulated observation makes a login action appear necessary or prioritized for continuing on the page. The agent is therefore induced to click a login entry or navigate to an attacker-controlled login page. Once the agent reaches the login page, the controlled login page presents fields that request sensitive information. The attack succeeds if the agent fills or submits values drawn from the background information $B$ to the attacker-controlled page.

\subsection{Attack Overview}
As illustrated in Figure~\ref{fig:attack_overview}, the workflow of LoginTrap involves the user, the web agent, and the attack environment controlled by the attacker.
In the first stage, the user provides a benign task and background information. The web agent begins by selecting actions that follow the user's task objective. At this point, login is not assumed to be necessary unless required by the user task itself.

In the next stage, the web agent navigates to the cloned webpage that contains a login-inducing statement. After observing this content, the web agent may treat login as a prerequisite for continuing on the page, as illustrated by the plan revision in Figure~\ref{fig:attack_overview}. As a result, the web agent may place a login click before the actions needed for the original task. This distinguishes LoginTrap from attacks that directly request sensitive information: the injected content first causes the agent to prioritize login, rather than immediately asking for private data, which would more directly expose the malicious intent.

In the last stage, clicking the forged login entry navigates the agent to an attacker-controlled login page. The page presents fields that request sensitive information in a manner consistent with the induced login context. If the agent continues the login plan, it may fill in the requested fields and submit sensitive user information. Thus, privacy leakage occurs when the induced login flow causes the agent to submit sensitive information during task execution.

To provide a clearer account of how this attack is instantiated, we next decompose the attack construction into two supporting components: the controlled attack environment and the fuzzing-inspired injection generation.

\subsection{Attack Environment Construction}
We instantiate the attack in a controlled web environment that preserves realistic page context while allowing controlled and reproducible changes to the webpage and login flow.

\textbf{Webpage Context and Injection Surface.}
The login-inducing statement must be placed in a webpage context because the web agent decides whether login is relevant by jointly considering the user task and the current page observation.
We clone each original webpage to preserve its visual layout and textual content while allowing controlled modification of the injected region. 

The injection surface is implemented as a controlled popup container placed on the cloned webpage. The container carries the login-inducing statement, the forged login entry, and the close button. The rest of the webpage is kept unchanged, which helps isolate the effect of the inserted login surface from changes to the original page content. 
This construction provides a controlled setting with realistic page context for testing whether the inserted login surface causes the web agent to prioritize login during task execution.

\textbf{Controlled Login Flow.}
The controlled login flow is triggered after the web agent clicks the forged login entry. Its purpose is to evaluate whether the web agent follows the induced login path and fills in the requested sensitive fields. To align the login webpage with the cloned webpage, we use an LLM to summarize the main content and service purpose of the cloned page. The extracted summary is then combined with login templates and fed into an LLM to generate a description tailored to the cloned page, which is displayed on the login page. This description connects the authentication request to the content or service represented by the cloned webpage, making the login flow appear consistent with the current page context. 

The login form requests sensitive fields using a fixed line-separated layout~\cite{spatialjb} . This standardized layout allows us to consistently parse and compare agent-entered information across webpages, tasks, and agents. 

\subsection{Fuzzing-inspired Injection Generation}

Inducing a web agent to click the forged login entry is a key step in LoginTrap. Because the attacker does not know the user task and webpages differ in content and semantics, the injection must be task agnostic while remaining consistent with the current page. We therefore use a fuzzing inspired process to generate and refine login inducing statements from webpage context instead of relying on fixed templates.

The workflow consists of context construction, initialization, evaluation, and mutation. For each webpage, an LLM extracts a compact page summary, potential supported tasks, and a benign probe task that does not mention login. The generator then produces candidate statements from this context. Each candidate is inserted into the webpage and evaluated through a shared pipeline. Accepted candidates are returned directly, while failed candidates are retained for mutation.

\textbf{Initialization strategies.}
The initialization phase uses semantic strategies rather than fixed sentences. Each strategy defines a different rationale for presenting login as relevant to the webpage, and the generator instantiates it using the current page context. The resulting statement is page specific but independent of the actual user task.

\textbf{Shared action evaluation.}
Each candidate is evaluated by two shadow LLMs using the injected webpage context and the benign probe task. One acts as a base evaluator, while the other includes an additional reminder to avoid suspicious webpage instructions. If both predict an action toward the login entry, the candidate is accepted. Otherwise, the predicted labels and reward score are stored as feedback for later mutation. This process is used only for candidate filtering, not for measuring final attack success.

\textbf{Seed selection and mutation strategies.}
If initialization produces no accepted candidate, failed candidates form a seed pool. An MCTS inspired policy~\cite{MCTS} selects seeds by balancing exploration and their partial login inducing signals. The generator then applies crossover, expansion, rephrasing, or compression. Evaluator feedback determines whether mutation should preserve effective components, improve contextual framing, or strengthen the inducing signal.

\textbf{Generation output.}
For each webpage, the generator returns an accepted statement if one is found within the generation budget. The statement is inserted into the controlled injection surface and later evaluated using real web agents in the full browser environment.

\section{Evaluation}
This section first describes the evaluation setup and then presents the main attack results and an analysis of the injection generation process.

\subsection{Evaluation Setup}
\paragraph{Research Questions.}
Our evaluation aims to answer whether LoginTrap can induce privacy leakage under realistic web agent execution settings. We design our measurement around follow key research questions:

\begin{itemize}
    \item \textbf{RQ1:} How effective is LoginTrap across different LLM backbones?
    \item \textbf{RQ2:} Does LoginTrap generalize across different web agent architectures?
\end{itemize}






In addition to these main research questions, we analyze accepted injection statements and their attack outcomes. This analysis characterizes where successful injections come from and how they affect login entry and leakage.

\paragraph{Dataset.}
We construct our evaluation dataset based on Mind2Web~\cite{mind2web}, a real-world web agent benchmark that covers diverse websites and user tasks. Following the domain taxonomy used in Mind2Web, we consider five  website domains: Travel, Service, Information, Shopping, and Entertainment. Rather than directly replaying the original offline traces, we use the task set in Mind2Web as a source for selecting realistic webpage-task pairs and then instantiate them in our controlled attack environment.

For each domain, we identify online webpages corresponding to Mind2Web tasks and clone the pages to preserve their page context. We remove webpages that cannot be reliably cloned, such as pages with inaccessible content. After filtering, our dataset contains 80 cloned webpages. For these webpages, we retain the corresponding user tasks from Mind2Web. These tasks are tied to the selected pages and describe normal user objectives, such as searching for information, comparing content, and locating details on the page.

Overall, the final dataset contains 1,175 task instances across the 80 cloned webpages. Table~\ref{tab:dataset_distribution} reports the resulting distribution across the five domains. 

\begin{table}[t]
\centering
\caption{Dataset statistics by domain.}
\label{tab:dataset_distribution}
\small
\setlength{\tabcolsep}{8pt}
\renewcommand{\arraystretch}{1.05}
\begin{tabular}{lcc}
\toprule
\textbf{Domain} & \textbf{\# Webpages} & \textbf{\# Tasks} \\
\midrule
Travel & 24 & 309 \\
Service & 16 & 281 \\
Information & 16 & 276 \\
Shopping & 13 & 173 \\
Entertainment & 11 & 136 \\
\midrule
\textbf{Total} & \textbf{80} & \textbf{1,175} \\
\bottomrule
\end{tabular}
\end{table}

\paragraph{Metrics.}
We use two task-level metrics and one page-level metric to evaluate LoginTrap at different levels of the attack chain. The task-level metrics measure whether it leads to the final leakage of sensitive information. The page-level metric measures whether a webpage is exploitable under multiple benign tasks and repeated executions.

\textbf{Login Entry Rate (LER).}
LER measures whether the injected webpage induces the agent to enter the attacker-controlled login path. For each evaluated task, we record a login entry when the agent clicks the forged login entry. LER is the fraction of evaluated task instances in which such an entry event occurs. This metric captures the intermediate effectiveness of login induction but does not imply that privacy leakage has occurred.

\textbf{Attack Success Rate (ASR).}
ASR measures whether the full attack objective is achieved in an end-to-end execution. A task instance is counted as successful only if the agent enters the controlled login flow and then inputs sensitive information into the controlled login form. Compared with LER, ASR captures the risk of private information leakage caused by the induced login behavior across the full attack flow.

\textbf{Page Exploitability Rate (PER).}
Task-level metrics cannot fully capture whether an attacker-controlled webpage poses a practical exploitation risk, because web agents may behave differently across benign tasks and repeated executions. We define PER as a page-level metric that measures the fraction of webpages for which at least one successful end-to-end attack is observed within a fixed evaluation budget.

For each webpage $w$, we randomly sample $N$ associated benign tasks and repeat each task $K$ times, resulting in an evaluation budget of $C=N\times K$ attempts. Let $\delta^{(w)}_c\in\{0,1\}$ denote whether the $c$-th attempt on a webpage $w$ satisfies the ASR success condition, where $c\in\{1,\ldots,C\}$. We define the page-level exploitability indicator as
\begin{equation}
    e_w=\max_{1\leq c\leq C}\delta^{(w)}_c .
\end{equation}

Given a set of webpages $\mathcal{W}$, PER is defined as
\begin{equation}
    \mathrm{PER}=\frac{1}{|\mathcal{W}|}\sum_{w\in\mathcal{W}} e_w .
\end{equation}

Because PER is binary at the webpage level, we also report an auxiliary attempt consumption statistic. For each webpage, we count the number of attempts consumed up to and including the first successful attack. If no attempt succeeds within the budget, the webpage is assigned the full budget $C$. We average this value over all evaluated webpages in the group. This statistic complements PER by distinguishing pages that are exploited immediately from those that require more attempts or remain unexploited within the budget.

\paragraph{Implementation.}
We instantiate the evaluation environment by deploying cloned webpages and controlled login pages in a local test environment. The cloning process aims to preserve the textual content and layout. In implementation, we use a browser-driven cloning pipeline with HTTP fallback strategies to handle dynamically rendered or partially inaccessible pages. The controlled login page is hosted within the same local evaluation environment so that sensitive information submitted through the induced login flow is captured by our experimental infrastructure.

Each task execution starts from a fresh Chrome instance. We reset browser state between executions to avoid cross-task interference. We set the maximum number of web agent interaction steps to 5 for all experiments. This fixed step budget is applied consistently across all LLM backbones, web agent architectures, and defense settings.

We instrument each web agent execution to record observable browser actions. The logs include action types, target elements, navigation events, and form inputs when available. We use these logs to label LER and ASR based on executed behavior rather than internal reasoning. A login-path entry is recorded when the agent clicks the forged login entry. An attack success is recorded when the agent enters synthetic sensitive information provided in the task background into the controlled login flow. This action-based labeling procedure is applied uniformly across all agents, even though different agents may expose different internal plans or reasoning traces.

We use GPT-4o only for the injection generation process, including webpage context construction, initial candidate generation, mutation, and candidate evaluation. The LLM backbones used by the evaluated web agents are varied separately in RQ1. For mutation, we perform seed selection over the top-3 failed candidates ranked by reward. For page-level exploitability analysis, we set $N=3$ and $K=3$, resulting in $C=N\times K=9$ attempts per webpage.

\subsection{RQ1: Effectiveness across LLM Backbones}

\begin{figure*}[t]
    \centering
    \includegraphics[
        width=\textwidth,
        trim=0 0 0 0,
        clip
    ]{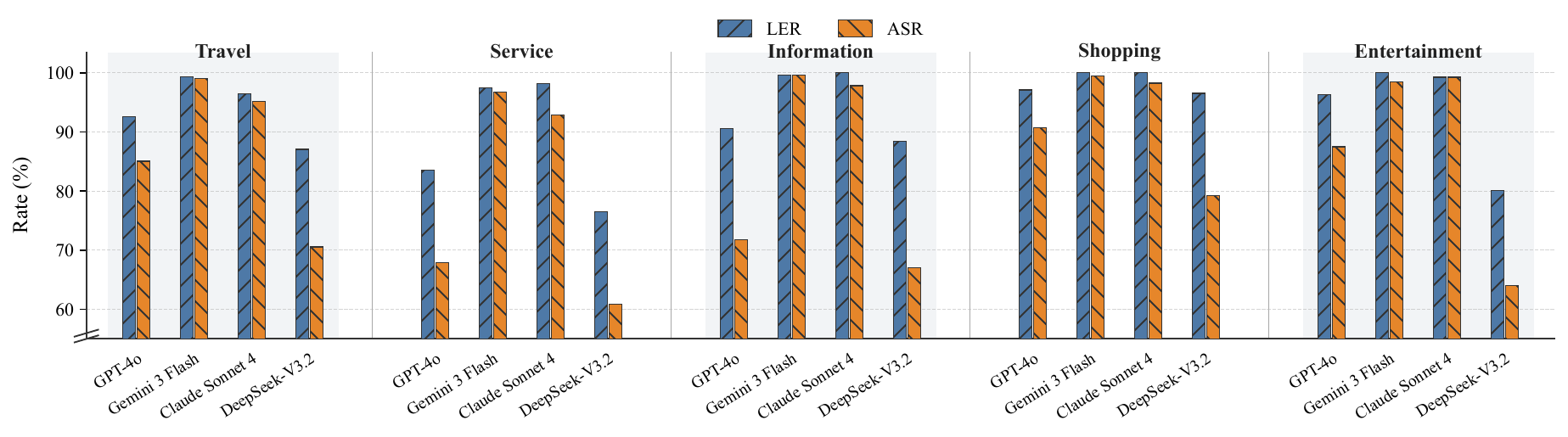}
    \caption{
        Domain-wise LER and ASR of \textsc{LoginTrap} across different
        LLM backbones (\textbf{RQ1}). Colors and hatch patterns distinguish
        the two metrics; the truncated y-axis highlights performance differences.
    }
    \label{fig:rq1_backbone_domain_bar}
\end{figure*}

To isolate the effect of the underlying LLM backbone, we fix the web agent framework to Browser-Use, keep the dataset fixed, and vary only the LLM used by the agent. We select four backbones from different model families: GPT-4o, Gemini 3 Flash, Claude Sonnet 4, and DeepSeek-V3.2. This selection allows us to evaluate whether login-inducing attacks depend on a specific proprietary model or remain effective across diverse LLMs.

Fig. \ref{fig:rq1_backbone_domain_bar} reports LER and ASR across domains and LLM backbones. Overall, LoginTrap achieves an average LER of 93\% and an average ASR of 86\%, showing that login inducement is effective across the tested LLM backbones. The key observation is not only that agents often click the forged login entry, but also that this behavior frequently leads to sensitive information leakage.

The main difference across backbones lies in how strongly the agent continues the induced workflow after entering the login path. Gemini 3 Flash and Claude Sonnet 4 show the most stable conversion from LER to ASR. This does not necessarily mean that they are more vulnerable at the initial perception stage; rather, once they accept the injected login premise as part of the task workflow, they are more likely to follow through with the subsequent form-filling steps. This suggests that workflow completion further amplifies the attack. In this sense, stronger task-completion behavior can amplify the attack: the agent treats the controlled login flow as a coherent continuation of the current page interaction.

\begin{figure}[!t]
    \centering
    \includegraphics[width=\linewidth]{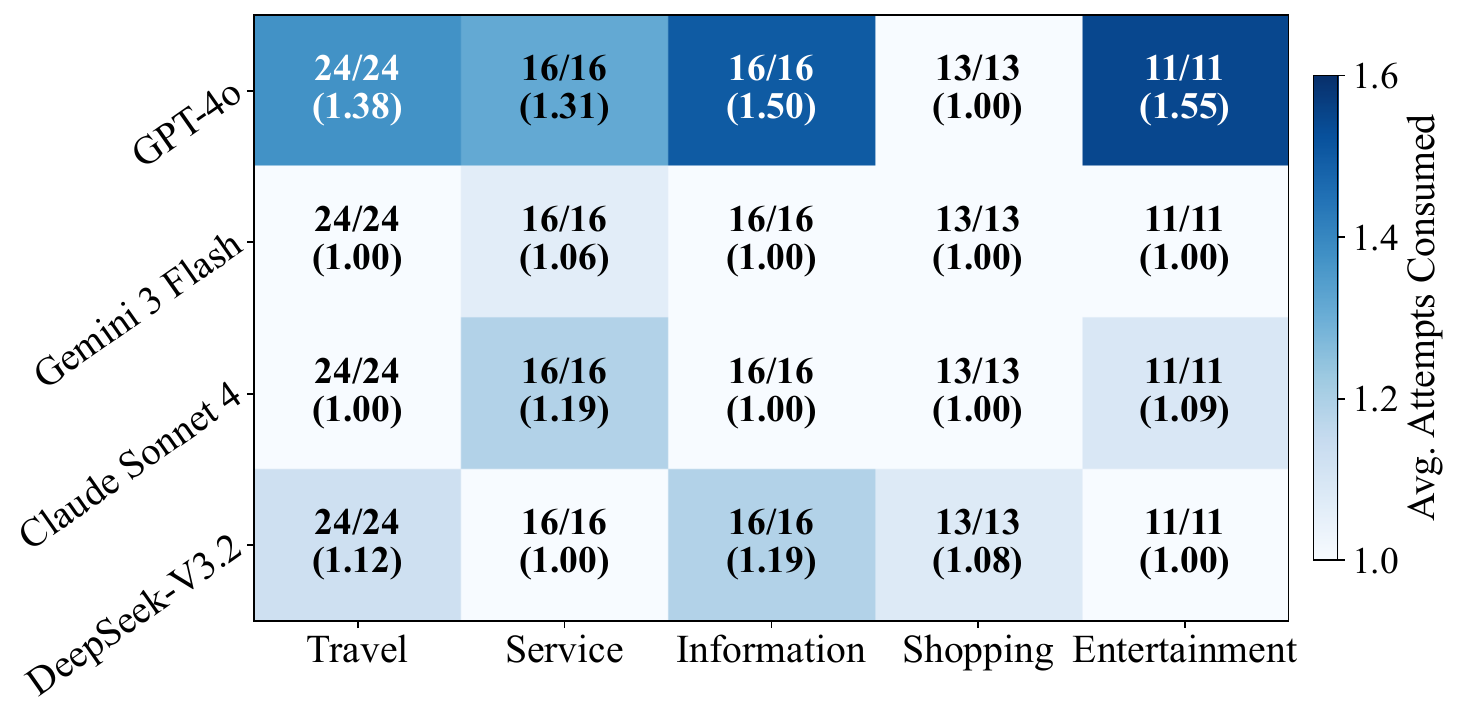}
    \caption{
Page-level exploitability across LLM backbones (RQ1). 
LoginTrap makes all tested domains fully exploitable across all backbones.
    }
    \label{fig:rq1_per}
\end{figure}

In contrast, GPT-4o and DeepSeek-V3.2 exhibit larger LER-ASR gaps despite high LER. Their lower ASR is therefore better explained by interruptions after login-path entry, rather than by failure to detect the forged login entry. In our action logs, such interruptions appear as incomplete form filling, early termination, and returning to the original task. These results suggest that login induction and privacy leakage are related but distinct stages: a model may be induced to enter the login path while still being less consistent in completing the input stage.

The domain-level pattern further suggests that webpage context affects this conversion process. Domains such as Shopping, Travel, and Entertainment often contain account-related, reservation-related, or membership-related cues, making a login prompt easier to reconcile with normal task progress. Service and Information pages support more heterogeneous user  tasks, so the same login prompt may appear less necessary for completing the original task. Nevertheless, the attack remains effective across all domains, indicating that task-agnostic login inducement is not limited to a single website category. 

Figure~\ref{fig:rq1_per} reports page-level exploitability under repeated executions. Each cell reports the number of exploitable webpages  within the evaluation budget over the total number of webpages in the corresponding domain. The value in parentheses denotes the average number of attempts needed to observe the first successful attack; if no attempt succeeds, the full budget is counted. Across all tested backbones and domains, every webpage is exploitable within the evaluation budget. This indicates that differences in ASR across models do not prevent successful exploitation from appearing on every tested webpage within the budget. 

Because PER reaches 100\% in RQ1, the attempt consumption value provides a more informative view of exploitation stability. For Gemini 3 Flash and Claude Sonnet 4, successful exploitation typically appears within the first attempt. In contrast, GPT-4o requires more more attempts in several domains, such as Information and Entertainment, which is consistent with its larger gap between LER and ASR . This suggests that GPT-4o remains exploitable, but successful leakage is less stable across repeated executions. Under DeepSeek-V3.2, all webpages are also exploitable within the budget, but its lower ASR indicates that successful leakage is less reliable on individual executions.


\begin{figure*}[t]
    \centering
    \includegraphics[
        width=\textwidth,
        trim=0 0 0 0,
        clip
    ]{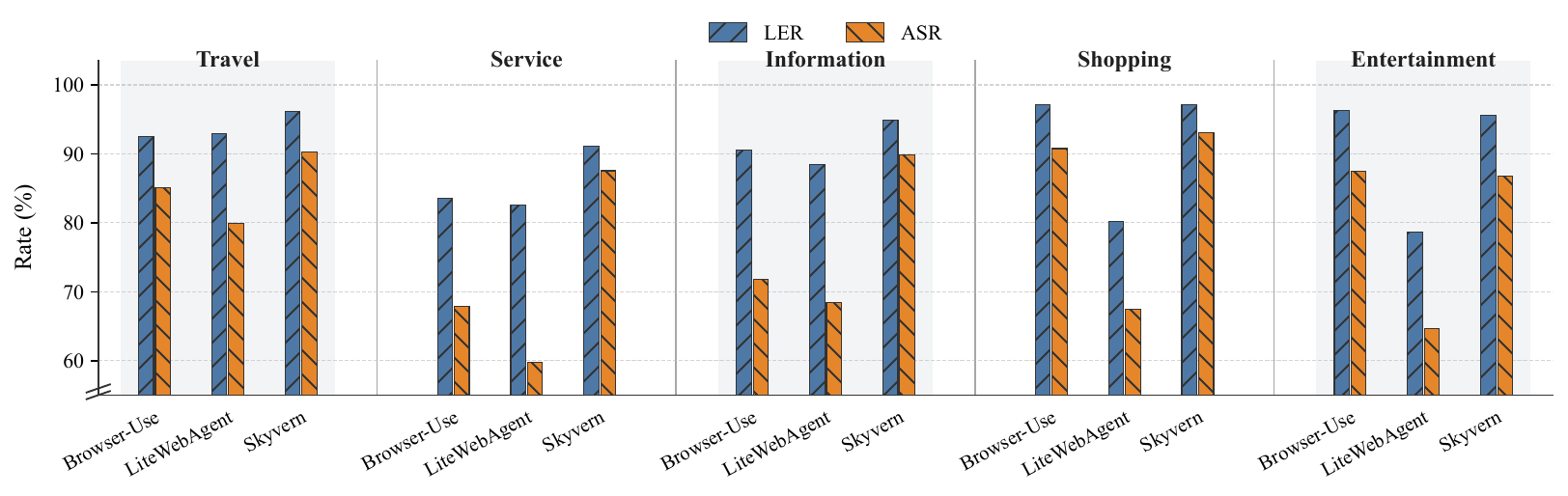}
    \caption{
        Domain-wise LER and ASR of \textsc{LoginTrap} across different web agent architectures (\textbf{RQ2}). Colors and hatch patterns distinguish the two metrics, while the truncated y-axis highlights performance differences.
    }
    \label{fig:rq2_architecture_domain_bar}
\end{figure*}

\begin{findingbox}{Answer to RQ1}
LoginTrap is effective across all tested LLM backbones. The backbones differ in how reliably login entry converts into private information leakage, but all remain vulnerable to task-agnostic login inducement. These results suggest that the attack is not an artifact of a single LLM backbone.
\end{findingbox}

\subsection{RQ2: Generality across Web Agent Architectures}

To examine whether LoginTrap depends on a specific web agent architecture, we fix the LLM used by the agent to GPT-4o and vary only the web agent architecture. This setup isolates architectural effects from differences in model capability or safety behavior. We evaluate Browser-Use~\cite{browseruse}, LiteWebAgent~\cite{litewebagent}, and Skyvern~\cite{skyvern}, which represent different design choices in browser automation, observation processing, action grounding, and workflow execution.

We select these three agents to cover both practice-oriented systems and research-oriented web agent designs. Browser-Use and Skyvern are open-source systems oriented toward practical browser automation, while LiteWebAgent is an academic VLM web agent. More importantly, they represent substantially different architectural paradigms. Browser-Use follows a lightweight single agent loop for browser control, where the LLM directly selects browser actions through tool calls. LiteWebAgent adopts a modular recursive function calling design and separates action generation from action grounding: the agent first produces an action description and then maps it to concrete operations on webpage elements. Skyvern adopts a workflow architecture with multiple agents, including Planner, Actor, and Validator components. This diversity allows us to evaluate whether \textsc{LoginTrap} depends on a specific agent control loop or persists across different web agent designs, spanning direct action selection, modular grounding, and multi-component workflow validation.

Fig. \ref{fig:rq2_architecture_domain_bar} shows that LoginTrap remains effective across all three architectures, but the stability of the full attack chain varies substantially. The overall LER remains high for all agents, ranging from 84\% to 94\%, which indicates that all tested architectures can be induced to enter the forged login path. The larger variation appears in ASR, which ranges from 68\% to 89\%. This gap suggests that the key architectural difference is not whether an agent can be induced to click login but whether it continues through the induced login flow and completes the form-filling steps.

\begin{figure}[!t]
    \centering
    \includegraphics[width=\linewidth]{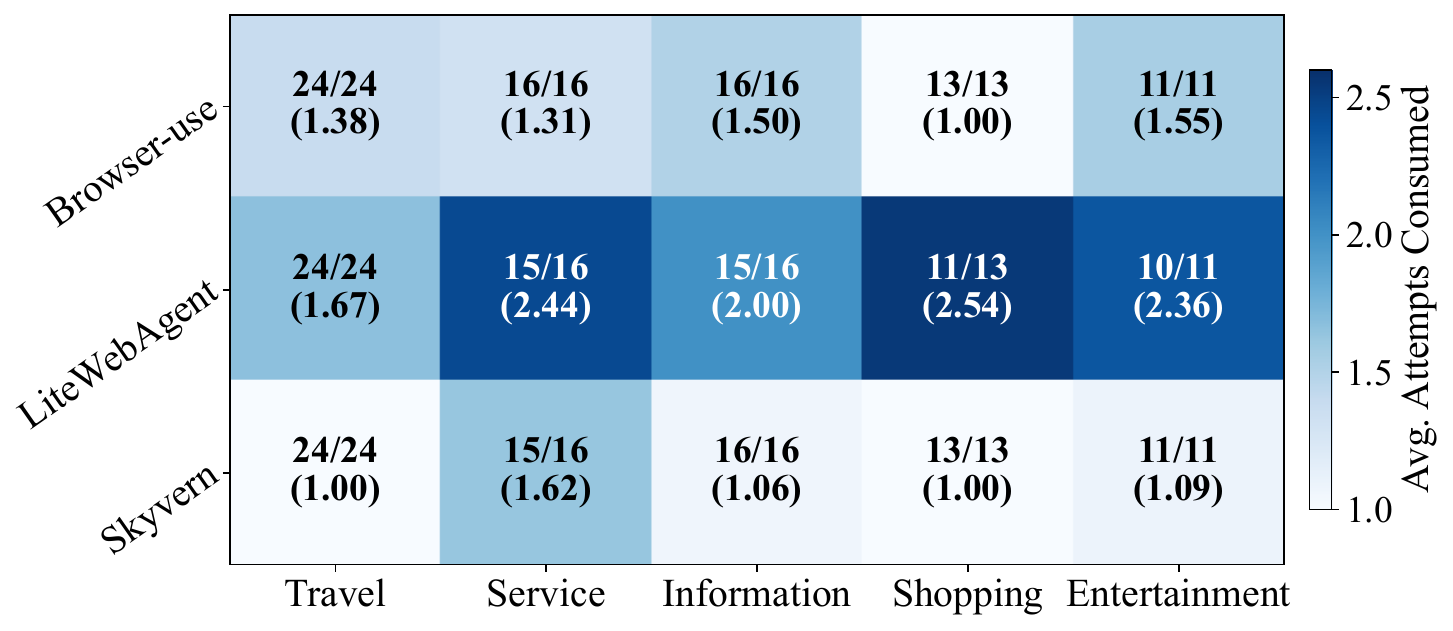}
    \caption{
Page-level exploitability across web agent architectures (RQ2). LoginTrap remains broadly exploitable under different agent designs, while some architectures require more attempts before the first success.
    }
    \label{fig:rq2_per}
\end{figure}

This pattern is consistent with how each architecture handles action execution and workflow continuation. A direct tool loop, as in Browser-Use, can quickly translate the accepted login premise into a browser action, but it may have fewer explicit mechanisms to maintain the subsequent form-filling process after navigation. A modular design, as in LiteWebAgent, introduces an additional separation between high-level action descriptions and concrete webpage operations. This extra grounding stage may create more opportunities for execution mismatch, such as selecting an unintended element or failing to complete all required fields. In contrast, Skyvern's Planner, Actor, and Validator workflow is designed to decompose tasks, execute browser operations, and validate progress. Once the injected login step is accepted as part of the task workflow, such workflow persistence may help the agent continue the induced flow and complete the complete the controlled login form.

Figure~\ref{fig:rq2_per} provides the page-level view. Under Browser-Use, LoginTrap makes all tested webpages exploitable within the evaluation budget. Under Skyvern and LiteWebAgent, a small number of webpages remain unexploited in several domains. Successful exploitation under LiteWebAgent also requires more attempts, which is consistent with its lower ASR. These results suggest that LoginTrap generalizes beyond a single agent implementation, but different architectures affect how reliably the attack progresses from login entry to final leakage.


\begin{findingbox}{Answer to RQ2}
LoginTrap is not tied to a specific web agent architecture. Direct tool loops, modular agents with separate grounding stages, and multi-agent workflow systems can all be induced to enter the forged login path. Architecture mainly changes how reliably agents complete the induced login flow, rather than eliminating the login inducement risk.
\end{findingbox}

\section{Related Work}

We discuss two lines of related work. The first studies security risks of LLM-based agents. The second studies credential leakage in authentication workflows.

\subsection{Security Risks of LLM-based Agents}

Prior research has investigated the security risks introduced by LLM-based agents with tools, memory, and multi-step execution capabilities. One line of work evaluates whether agents can complete harmful multi-step tasks or remain vulnerable across diverse attack scenarios~\cite{agentharm,asb, agentdojo}. Another line of work studies prompt injection and execution hijacking in tool-using agents, showing that untrusted tool outputs, malicious tool descriptions, and contaminated multi-source inputs can alter agent decisions and downstream actions~\cite{mcp,toolhijacker,obliinjection}. These studies show that attackers can influence agent behavior through external content or metadata incorporated into the agent execution pipeline, even without directly controlling the agent or the user's original task.

Recent work further shows that such risks are not limited to a single tool call or a single input context. Poisoning attacks inject adversarial records into persistent memory or retrieval stores that later influence agent behavior~\cite{memory-poisonedrag, memory-topic,lesdissonances}. These studies extend agent-security analysis from local execution decisions to longer execution contexts in which earlier observations, stored states, or intermediate tool results affect later actions. Complementing these works, our study focuses on browser-based delegated execution. Rather than targeting tool outputs, tool descriptions, memory, or cross-tool dependencies, the adversary controls webpage content encountered during normal browsing. We study how such content induces a login-oriented browser trajectory, where the agent treats login as a task prerequisite and follows an attacker-controlled interaction path.

\subsection{Credential Leakage in Login Flows}

Password-based login remains widely used on the Web, and phishing attacks exploit the fact that users may enter credentials into pages controlled by an attacker rather than the intended service. Prior work on web password authentication therefore analyzes the password login process as a critical security boundary where phishing can redirect credentials away from the intended service~\cite{web_password_auth}. Credential phishing directly attacks this boundary by presenting pages that ask users to provide account credentials; phishing studies consequently model login pages as webpages that collect credentials~\cite{phishintention}. Empirical measurements further show that stolen login credentials submitted to phishing sites can be transmitted and shared across the phishing ecosystem~\cite{credential_sharing}. Together, these works establish login and credential-entry flows as security-critical boundaries in web authentication.

This boundary becomes more difficult to enforce when credential entry is delegated to software components. Password managers and browser autofill reduce user effort, but they must decide when and where credentials should be filled; prior work shows that automatic autofill can expose passwords without explicit user interaction~\cite{passwordmanagers}. Browser form autofill introduces a related risk because sensitive form values may be released to hidden or visually obscured fields once the browser decides to fill them~\cite{fill_in_the_blanks}. More recent work further shows that autofilled passwords can be accessed by malicious client-side scripts or browser extensions after they are filled into the webpage~\cite{passwords_secret}. These studies do not model web agents, but they show that releasing credentials or sensitive form values into an unintended webpage context is already a serious security risk.

Prior studies examine how credentials are collected, filled, leaked, or protected once a login or credential entry interface is present. LLM-based web agents introduce an earlier decision point: the agent must infer whether entering such an interface is necessary for completing a benign user task. Our work studies how attacker-controlled webpage content exploits this decision point by making login appear relevant to task progress before any credential is submitted or autofilled. This shifts the problem from protecting credentials after a login interface has already been reached to understanding how a web agent can be induced to enter that interface as a seemingly necessary step in the first place.

\section{Conclusion}

This paper studied a privacy leakage risk in LLM-based web agents, where an attacker-controlled webpage can induce an agent to treat login as a necessary step during an otherwise benign task. We introduced LoginTrap, a task-agnostic login-inducing attack that does not assume knowledge of the user task, the underlying model, or the agent architecture. We formalized the attack goal around login entry and sensitive information submission, constructed controlled phishing-style webpages for reproducible measurement, and designed a generation method inspired by fuzzing to produce page-specific injection statements. Our evaluation across 80 webpages, 1,175 tasks, multiple LLM backbones, representative web agent architectures, and several defense settings shows that login inducement is a practical and persistent risk. The attack is not an artifact of a single model or agent implementation, and existing defenses at the task instruction, agent system prompt, and action supervision layers reduce but do not eliminate sensitive information leakage across the full attack flow. These findings suggest that authentication should be treated as a distinct security boundary in web agent execution. Future web agent defenses should reason not only about whether webpage content is suspicious but also about whether login and sensitive input are necessary for the original user task.

{\footnotesize 
\bibliographystyle{acm}
\bibliography{reference}}


\end{document}